# Observational evidence for the associated formation of blobs and raining inflows in the solar Corona

E. Sanchez-Diaz[1,2]: 9 avenue Colonel Roche, BP 44346 - 31028 Toulouse Cedex 4A (France); eduardo.sanchez-diaz@irap.omp.eu

A. P. Rouillard[1,2]: 9 avenue Colonel Roche, BP 44346 - 31028 Toulouse Cedex 4A (France); alexis.rouillard@irap.omp.eu

J. A. Davies[3]: Science and Technology Facilities Council Rutherford Appleton Laboratory Harwell Campus Didcot OX11 0QX (United Kingdom); jackie.davies@stfc.ac.uk

B. Lavraud[1,2]: 9 avenue Colonel Roche, BP 44346 - 31028 Toulouse Cedex 4A (France); benoit.lavraud@irap.omp.eu

N. R. Sheeley[4]: Naval Research Laboratory Code 7600 4555 Overlook Ave SW Washington, DC 20375 (United States); neil.sheeley@nrl.navy.mil

R. F. Pinto[1,2]: 9 avenue Colonel Roche, BP 44346 - 31028 Toulouse Cedex 4A (France); rui.pinto@irap.omp.eu

E. Kilpua[5]: Department of Physics P.O. Box 64 FI-00014 University of Helsinki, Helsinki (Finland); emilia.kilpua@helsinki.fi

I. Plotnikov[1,2]: 9 avenue Colonel Roche, BP 44346 - 31028 Toulouse Cedex 4A (France); illya.plotnikov@irap.omp.eu

V. Genot[1,2]: 9 avenue Colonel Roche, BP 44346 - 31028 Toulouse Cedex 4A (France); vincent.genot@irap.omp.eu

[1]Institut de Recherche en Astrophysique et Planétologie, Paul Sabatier University, Toulouse, France
[2]Centre National de la Recherche Scientifique, UMR 5277, Toulouse, France
[3]RAL Space, STFC-Rutherford Appleton Laboratory, Harwell Campus, UK
[4]Space Science Division, Naval Research Laboratory, Washington, DC, USA
[5]Space Physics Department, University of Helsinki, Finland

# ABSTRACT

The origin of the slow solar wind is still a topic of much debate. The continual emergence of small transient structures from helmet streamers is thought to constitute one of the main sources of the slow wind. Determining the height at which these transients are released is an important factor in determining the conditions under which the slow solar wind forms. To this end, we have carried out a multipoint analysis of small transient structures released from a north-south tilted helmet streamer into the slow solar wind over a broad range of position angles during Carrington Rotation 2137. Combining the remote-sensing observations taken by the Solar-TErrestrial RElations Observatory (STEREO) mission with coronagraphic observations from the Solar and Heliospheric Observatory (SoHO) spacecraft, we show that the release of such small transient structures (often called blobs), which subsequently move away from the Sun, is associated with the concomitant formation of transient structures collapsing back towards the Sun; the latter have been referred to by previous authors as 'raining inflows'. This is the first direct association between outflowing blobs and raining inflows, which locates the formation of blobs above the helmet streamers and gives strong support that the blobs are released by magnetic reconnection.

# 1. INTRODUCTION

An important contribution to the slow solar wind, and to its variability, is thought to come from small density features, which continually emerge from the inferred source regions of the slow solar wind (Sheeley et al. 1997). The term "blob" is used to define such *small-scale, outward-moving density enhancements*, in particular when detected in white-light by coronagraphs (Sheeley et al. 1997) and heliospheric imagers (e.g. Rouillard et al. 2010a). At heliospheric altitudes, blobs are best observed when the fast solar wind is radially aligned behind the slow solar wind (Rouillard et al. 2008). The resulting compression leads to the formation of a Stream Interaction Region (SIR), or Corotating Interaction Region (CIR) if it survives for at least one solar rotation (Jian et al. 2006). This compression allows for clearer observation of the pre-existing small-scale structuring of the slow solar wind. Transient density structures are also detected in the solar wind on other spatial and temporal scales (Viall et al. 2009, 2010, 2015, Kepko et al. 2016).

Observations of blobs from multiple vantage points, by the STEREO coronagraphs, show that they are the signatures of small coronal loops with the structure of flux ropes (Sheeley et al. 2009). STEREO observations have also enabled their tracking to 1 AU; the in-situ signatures of these blobs are reminiscent of closed magnetic field lines and flux ropes (Rouillard et al. 2010a,b). Note that similar transient structures had previously been detected in in-situ measurements alone (e.g. Crooker et al. 1996, 2004; Kilpua et al. 2009). Smaller-scale periodic density structures have recently been detected in high-resolution in-situ measurements (Kepko et al. 2016) that could relate to the density structures released into the solar wind at higher frequencies (Viall et al. 2009, 2010, 2015). 70-80% of in-situ measurements of the slow solar wind revealed such structuring (Viall et al. 2009).

The formation of blobs has, thus, been argued to imply magnetic reconnection in the solar corona (e. g. Lionello et al. 2005). Such magnetic reconnection should produce two density enhancements that separate in space from one another over time. This hypothesis is supported by the separate imaging of periodic inward-moving density enhancements, known as raining inflows (Sheeley & Wang 2001, 2002, 2007, 2014; Sheeley et al. 2001; Wang et al.

1999). Nevertheless, as yet there has been no direct observational evidence of the association between blobs released outwards and raining inflows. This is thought to be due to instrumental limitations (see section 2).

In this manuscript, we take advantage of the coronal magnetic field topology during Carrington Rotation 2137, accompanied by an optimal configuration of the STEREO and L1 spacecraft (see section 3), to validate the hypothesis of a common physical mechanism forming blobs and raining inflows.

## 2. OBSERVATIONAL CONSTRAINTS

The LASCO instrument package (Brueckner et al. 1995), on-board the SOHO spacecraft (Domingo et al. 1995) comprises three coronagraphs that observe white light that has been Thomson-scattered, proportional to the plasma density ─ C1 with a field of view (FOV) extending from 1.1 to 3 solar radii ($R_s$), and which operated only from 1996 to 1998, C2, with a FOV from 1.5 to 6 $R_s$, and C3, the outermost one, with a FOV from 3.7 to 32 $R_s$.

C2 has observed a large number of inflows during solar cycle 23 (Sheeley & Wang 2007, 2014). Raining inflows, one category of such inflows, are defined as a multitude of small density structures that fall back, periodically, towards the Sun at all latitudes when a highly tilted coronal current sheet or neutral line passes in the plane of sky (POS). Only 2% of inflows of all categories are observed to be associated with an outflowing component. Such inflows are generally associated with coronal mass ejections (Sheeley & Wang 2002), although some isolated inflows, known as either falling columns or falling curtains (depending on their angular size), have been observed in conjunction with corresponding outflows (Sheeley & Wang 2007). No obvious outflowing component has ever been clearly associated with the raining inflows (Sheeley & Wang 2007, 2014). Either there was no corresponding outflowing component or its brightness fell below the detection threshold of C3. In this paper, we make use of the instrumentation offered by the SECCHI package (Howard et al. 2008), which provides complementary white-light observations on board the twin STEREO spacecraft (Kaiser et al. 2008), to search for a potential outflowing component. The SECCHI package on each STEREO spacecraft consists of an inner coronagraph (COR1) that observes the corona between 1.4 and 4 $R_s$ (POS), an outer coronagraph (COR2) that observes between 2.5 and 15 $R_s$ and the Heliospheric Imager (HI), which observes from the outer edge of the COR2 field of view (FOV) out to 1 AU and beyond. DeForest et al. (2014) have used COR2 images to detect inward propagating waves that the authors suggest are consistent with reconnection inflows (Tenerani et al. 2016).

Each HI instrument contains two wide-angle cameras. The FOV of the inner HI (HI1) camera extends from 4° to 24° elongation along the ecliptic and the FOV of the outer HI (HI2) camera, from 19° to 89° elongation. HI has been successfully used to track blobs over an extended range of heliographic longitudes, and in particular blobs that have been compressed by CIRs. Such compression counteracts, at least in part, the strong radial expansion experienced by these structures as they propagate outward (Rouillard et al. 2008; Sheeley et al. 2008).

The inflows that have been observed with C2 typically form at 5 $R_s$. Hence, COR2 (FOV: 2.5-15 $R_s$) should be ideally suited to observe both the inflow and any associated outflow. The heights imaged by HI1 (from 15 $R_s$ POS), which are beyond the typical Alfvén radius

(Goezler et al. 2014), are well beyond those imaged by the C2 coronagraph (out to 6 $R_s$). For this reason, no inflows are observed in the HI1 FOV. COR2 images the gap between the C2 and HI1 FOVs starting at lower heights than C3 (3.7 $R_s$), where blobs are still dense enough to be clearly observed with a coronagraph. C2 is better suited than COR2 to observe inflows because the latter typically form at mid heights in the C2 FOV. Therefore the inflowing motion can be captured in consecutive images.

Over a solar cycle timescale, the total number of inflows of all kinds is well correlated with solar activity (Sheeley & Wang 2014). This explains why few raining inflows have been reported with STEREO, as the mission has operated thus far during the least active solar cycle of the space era. On timescale of month, the total inflow rate is better correlated with the gradient of the non-axisymmetric quadrupolar component of the coronal magnetic field (Sheeley & Wang 2014). The solar corona achieved a topology such that this component was important during Carrington rotation 2137 (May to June 2013), with a highly tilted neutral line, making it a favorable period to observe raining inflows.

## 3. OVERVIEW OF THE PERIOD OF STUDY

Figure 1a shows an extreme ultraviolet (EUV) image of the solar corona. The dark region near disk center, indicative of cold plasma, is a large coronal hole that passed through the central meridian of the Sun on 2013 May 29. This corresponds to a region of open magnetic filed lines, all of them with the same magnetic polarity. The presence of this extensive structure forced a strong excursion of the neutral line, which forms at higher altitude, to a north-south orientation as revealed by a Potential Field Source Surface (PFSS) extrapolation of photospheric magnetic fields (Figure 1b). The PFSS extrapolation (Schrijver & DeRosa 2003) is based on evolving surface magnetic maps into which are assimilated data from the Helioseismic and Magnetic Imager (HMI; Scherrer et al. 2012) onboard the Solar Dynamics Observatory (SDO).

The orbital configuration of STEREO and SOHO during this period is presented in the polar view of the ecliptic plane depicted in Figure 1c. The STEREO spacecraft were 140° ahead of and behind the Earth and separated by 80° from each other. The location of a CIR that passed through the FOV of the HI instruments, and will be presented later, is shown in Figure 1c as a blue spiral. We will demonstrate that this CIR was induced by the giant coronal hole shown in Figure 1a. Over several days, the CIR, the coronal hole and the north-south oriented neutral line corotated from the longitude of Earth to that of STEREO-A. Hence, they passed through the FOVs of LASCO and SECCHI, giving us the opportunity to study the variability of the corona and the solar wind from multiple vantage points and with different instruments. The CIR was detected by the in-situ instruments on board the STEREO and L1 spacecraft during several Carrington rotations.

Figure 2 presents a sequence of running-difference images derived from remote-sensing instruments onboard STEREO-A and SOHO. They are constructed by subtracting the preceding image from the current one. Such processing highlights propagating density enhancements, which are manifest as a bright followed by a dark feature in their direction of motion. The coronagraphs and HI instrument detected multiple blobs over the broad range of latitudes spanned by the north-south oriented neutral line. The structures imaged by HI1 have loop-like aspects, but are far more numerous over all observed latitudes than seen in previous studies (e.g. Rouillard et al. 2011). The loop-like structures appear to form larger-scale poly-lobed arcs in the HI1-A images extending over a broad range of latitudes. One such poly-lobed arc is indicated by a series of white arrows. Analysis of the variability along an east-west oriented neutral sheet (i.e. as a function of longitude) is difficult with images taken from the ecliptic plane because white-light features are integrated along the line of sight. By contrast, images of a north-south oriented neutral sheet allow us to study, at a single point in time, the distribution of streamers blobs over an extended surface area of the neutral sheet.

In the coronagraph images, we can observe these same structures during the earlier phase of formation. The second column of Figure 2 shows an example of one of the blobs observed by COR2. This blob is associated with an inflow, also imaged by COR2. When the neutral line passes into the POS of SOHO, C2 observes some raining inflows in addition to the blobs. Figure 2 (left) shows a raining inflow as a density depletion (a dark region leading a bright region in a difference image) observed by C2. Inflows can be observed as either density depletions or enhancements (Wang et al., 2007).

## 4. OBSERVATIONS

Figure 3 shows six time-elongation maps (J-maps; e.g. Davies et al. 2009) constructed using HI1/2-A running-difference data at three different Position Angles (PAs) during the passage of the CIR of interest, from 2013 May 28 to June 06. The tracks visible on these maps are the signatures of blobs.

We identified some of the separate tracks in each J-map as blobs and assigned each a theoretical trajectory, which is overplotted as a red solid line in the right hand panels. This theoretical trajectory is calculated assuming a constant velocity and direction of propagation. The computed velocity is 400 km s$^{-1}$, which is the velocity of the slow wind part of the CIR measured in-situ. The direction of each blob is determined by the Carrington longitude of the neutral line at the time of its emergence, given by the PFSS extrapolation of the photospheric magnetic field (Figure 1b). The initial condition is chosen such that the theoretical trajectory and the corresponding trace cross, at the same time, an arbitrary height of 20 $R_s$. This trajectory is projected into the J-map using Equation 1 of Rouillard et al. (2008). Most of the theoretical trajectories are consistent with the actual observed traces. This is qualitatively in agreement with the source of the blobs being close to the neutral line (Plotnikov et al. 2016).

Since the blobs are already visible at low coronal altitudes (Sheeley et al. 1997), some of the traces observed by HI likely start in the coronagraph FOVs. In order to track them down to their origin, we combined the J-maps derived from COR2 and the lower part of HI1. Figure 4 shows STEREO-A J-maps from 125° (upper panel) and 80° PA (lower panel) that combine COR2-A data with observations from the near-Sun part of the HI1-A FOV. The J-maps cover the estimated time of passage of the neutral line through the left limb of the STEREO-A POS, from 2013 May 29 to June 01. Many of the blobs seen by HI1-A during this period can be traced back to inflows seen by COR2-A, and every inflow is associated with a blob observed first by COR2-A and then by HI1-A. The first track indicated by a sequence of three arrows in the top panel of Figure 4 corresponds to the inflow shown in Figure 2 (middle column). The separation between the blobs and the inflows is observed between 5 and 6 $R_s$ over the center of the Sun.

The height of this separation makes COR2 much better suited than C3 to observe it. The separation occurs close to the inner edge of the C3 FOV, where coronagraphs have most visibility issues, and close to the middle of COR2 FOV. It would not be possible to observe this separation in C2 because, at most, one or two frames of the outflowing part would be available, and these would be at the outer edge of the C2 FOV where the signal to noise ratio is low. Even if the outflow reached the inner edge of the C3 FOV, we could not observe it due to a lower sensitivity of C3 compared with C2 and COR2.

If the raining inflows, traditionally observed by C2, are related to the outflowing blobs observed by HI-A, originating at the neutral line, we would expect to observe raining inflows when the neutral line is at the POS of SOHO. According to the PFSS extrapolation, the neutral line is at the SOHO POS between 3 and 4 June, while some blobs are still being observed by HI-A.

Figure 5 presents three J-maps derived from LASCO C2 running-difference data at three different PAs around the estimated time of passage of the neutral line through the right limb of the POS of SOHO (2013 June 01 to 05). Raining inflows are seen at all PAs at a rate of 3 to 5 inflows per day or one about every 6 to 8 hours. As usual, the raining inflows are not clearly associated to discernable outflows in C2 data, or in C3 data, but their periodicity is similar to that of the blobs clearly detected at higher altitudes by HI during the same period of time and in previous studies (e.g. Rouillard et al. 2010a). Similarly to the inflows observed with COR, they typically approach the Sun with average velocities around 100km s$^{-1}$, accelerating at their early life and decelerating below 4 $R_s$.

## 5. DISCUSSION

Some of the blobs observed by HI were also observed by COR2. J-maps constructed using COR2 show the first clear link between these blobs and raining inflows (Figure 4). The separation between the outflowing blob and the raining inflow was observed around 5-6 $R_s$, well above the location of helmet streamers. This does not preclude the possibility of inflow-outflow separations happening at other heights, in particular close to or below the inner edge of the C2 FOV (1.5 $R_s$). No additional separations occurring between 3 and 4 $R_s$ are observed during this period; these heights are close to the poorly imaged inner edge of COR2 but are situated near the middle of the C2 FOV, where sensitivity is highest and inflows would be detected if they occurred.

These inflows are clearly observed by COR2 when the coronagraph observes the streamer, and its associated neutral line, passing through its POS. The inflows stop being observed when the neutral line rotates out of the POS, due to the limited sensitivity of coronagraphs away from the POS. By letting that streamer rotate into the SOHO POS, we see many more inflows in the C2 images than in COR2 images, which fact confirms the corotation of raining inflows (Figure 5). A close association had been already found between raining inflows and the coronal neutral line (Sheeley et al. 2001). Here we link these dynamical processes with the continual outflow of transient structures typically measured in the solar wind with HI. The co-existence near 5-6 $R_s$ of oppositely directed magnetic field and the bursty nature of outflow/inflow occurrence supports the idea that magnetic reconnection is the key mechanism for the formation of blobs. The observation of raining inflows by LASCO followed by the detection of inflow-outflow pairs by COR2 shows that raining inflows are associated with in/out pairs and that this mechanism generates a continual stream of (1) outward-moving blobs in the form of loops and twisted magnetic fields and (2) inwardly-collapsing magnetic loops.

It has been both argued that the entire slow solar wind might be made of a continuous release of transients (e.g. Einaudi et al. 2001; Lapenta and Knoll, 2005; Antiochos et al. 2011) or may be composed of two components, one transient and one resulting from a continual wind flow heated and accelerated for instance by wave-particle interactions at highly expanded flux tubes (e.g. Wang et al. 2009). In-situ measurements also suggest two different sources of slow solar wind (Kasper et al. 2007, Stakhiv et al. 2015, 2016). Our results suggest that a transient part is formed above the helmet streamers as a consequence of magnetic reconnection. Ongoing work combining images with in-situ measurements at 1 AU aims to quantify the contribution of this transient part to the slow wind and explore their magnetic connectivity to the Corona. We note that Solar Probe Plus and Solar Orbiter will soon obtain unprecedented imaging of the tips of streamers providing the observations necessary to address these questions in detail. Magnetic reconnection is a ubiquitous mechanism in space and astrophysical plasmas. The same mechanism that explains magnetic reconnection during bursty bulk flows in the Earth's magnetotail has been proposed to occur in the solar corona (Birn & Hesse 2009; see also Linton & Moldwin 2009). Here, we find strong evidence that magnetic reconnection is taking place at small temporal and spatial scales at high coronal altitudes (and not only in association with pulses of Poynting flux in solar flares), thus providing further evidence for an analogy to the recurring substorm process at Earth.

## 6. CONCLUSION

A common origin for raining inflows and outflowing blobs had been previously suggested (e.g. Wang et al. 2000; Lionello et al. 2005). However, previous studies have observed these transients separately. Here, a very highly tilted neutral line that rotated into the STEREO POS have allowed us to observe for the first time raining inflows separating from blobs at 5-6 $R_s$ yielding the observational evidence that both raining inflows and blobs are products of the same process. Solar Probe Plus, during its closest approach (9 $R_s$) will fly very close to the observed point of inflow-outflow separation providing some additional insights on the formation mechanisms of these transients.

## Acknowledgements

We acknowledge usage of the tools made available by the plasma physics data center (Centre de Données de la Physique des Plasmas; CDPP; http://cdpp.eu/), the Virtual Solar Observatory (VSO; http://sdac.virtualsolar.org), the Multi Experiment Data & Operation Center (MEDOC; https://idoc.ias.u-psud.fr), the French space agency (Centre National des Etudes Spatiales; CNES; https://cnes.fr/fr) and the space weather team in Toulouse (Solar-Terrestrial Observations and Modelling Service; STORMS; https://stormsweb.irap.omp.eu/). This includes the data mining tools AMDA (http://amda.cdpp.eu/) and CLWEB (clweb.cesr.fr/) and the propagation tool (http://propagationtool.cdpp.eu). RFP, IP, ESD acknowledge financial support from the HELCATS project under the FP7 EU contract number 606692. NRS acknowledges NASA and CNR for financial support. The STEREO SECCHI data are produced by a consortium of RAL (UK), NRL (USA), LMSAL (USA), GSFC (USA), MPS (Germany), CSL (Belgium), IOTA (France) and IAS (France). The ACE data were obtained from the ACE science center. The WIND data were obtained from the Space Physics Data Facility

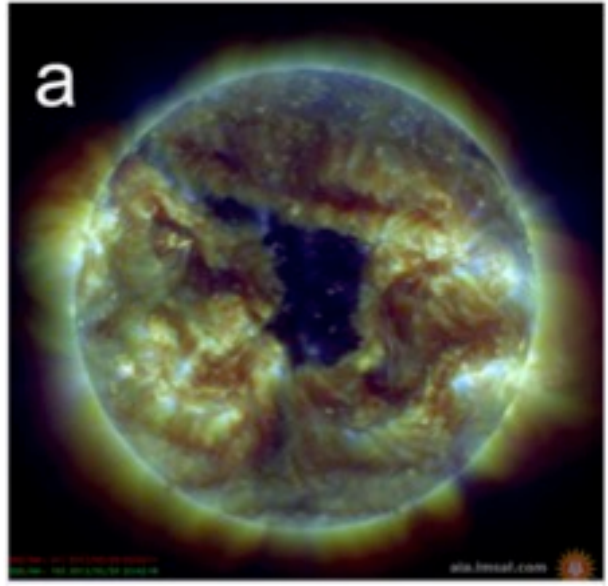
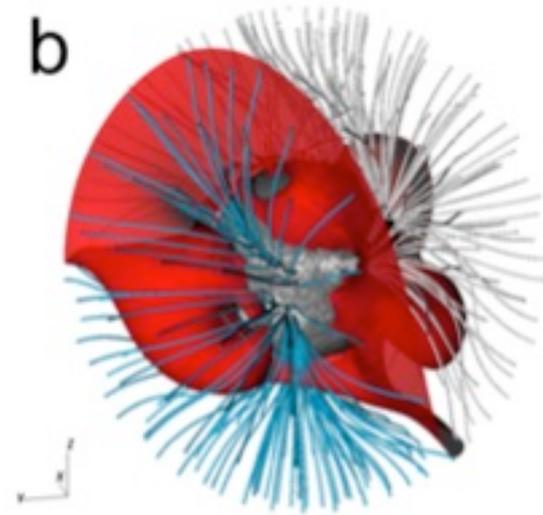
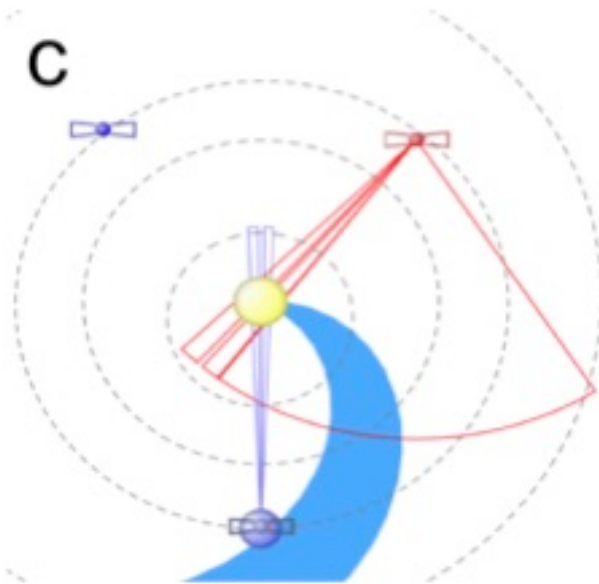

**Figure 1: Top:** AIA image of the solar corona (193 and 211 Å) on 2013 May 29. **Middle:** PFSS reconstruction of the coronal magnetic field for solar rotation 2137 with the neutral sheet (red sheet). Each line color corresponds to a different polarity of the magnetic field lines. **Bottom:** polar map of the ecliptic plane with the position of Earth (blue dot, not to scale), Sun (yellow dot, not to scale), STEREO-A (red butterfly, not to scale) and B (blue butterfly, not to scale), the FOV of STEREO HI-A (red triangle), COR2-A (red double triangle) and C2 (blue double triangle) and the CIR (blue spiral) on 2013 June 03 with an arbitrary width. Plot generated the propagation tool.

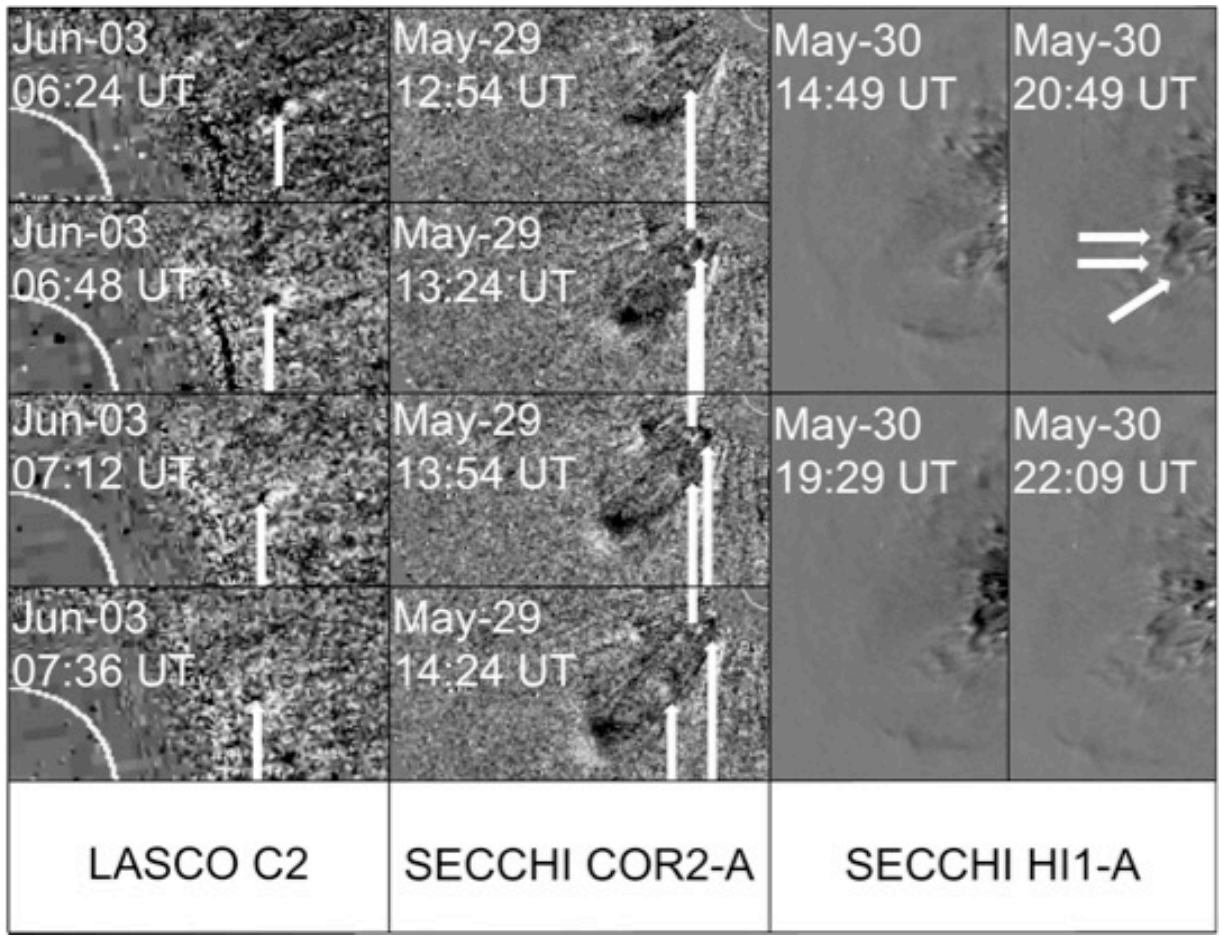

**Figure 2:** Sequence of images of a portion of the LASCO C2 FOV from 2013 June 03 06:24:05 UT to 07:36:05 UT (left), SECCHI COR2-A from 2013 May 30 12:54:00 UT to 14:24:00 UT (middle) and SECCHI HI1-A from 2013 May 30 14:49:01 UT to 22:09:01 UT (right) The white arrows show a raining inflow (left), an in/out pair (middle) and a poly-lobed arc (right) as examples.

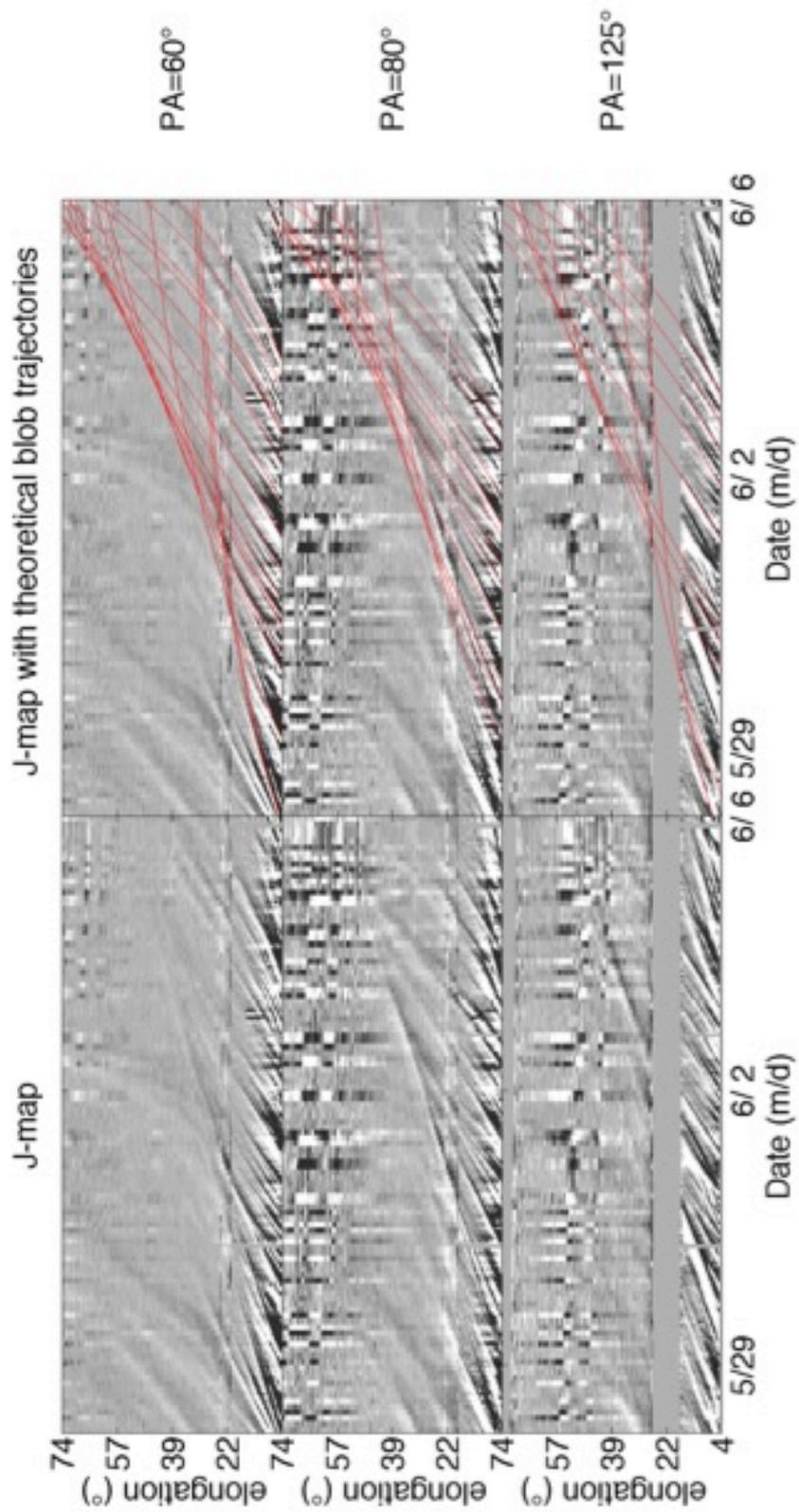

**Figure 3:** Left: J-maps of HI-A at 125° (top), 80° (medium) and 60° (bottom) position angle during the passage of the CIR, from 2013 May 28 to June 06. Right: copies of the same J-maps as left with the theoretical blob trajectories overplotted (red solid lines).

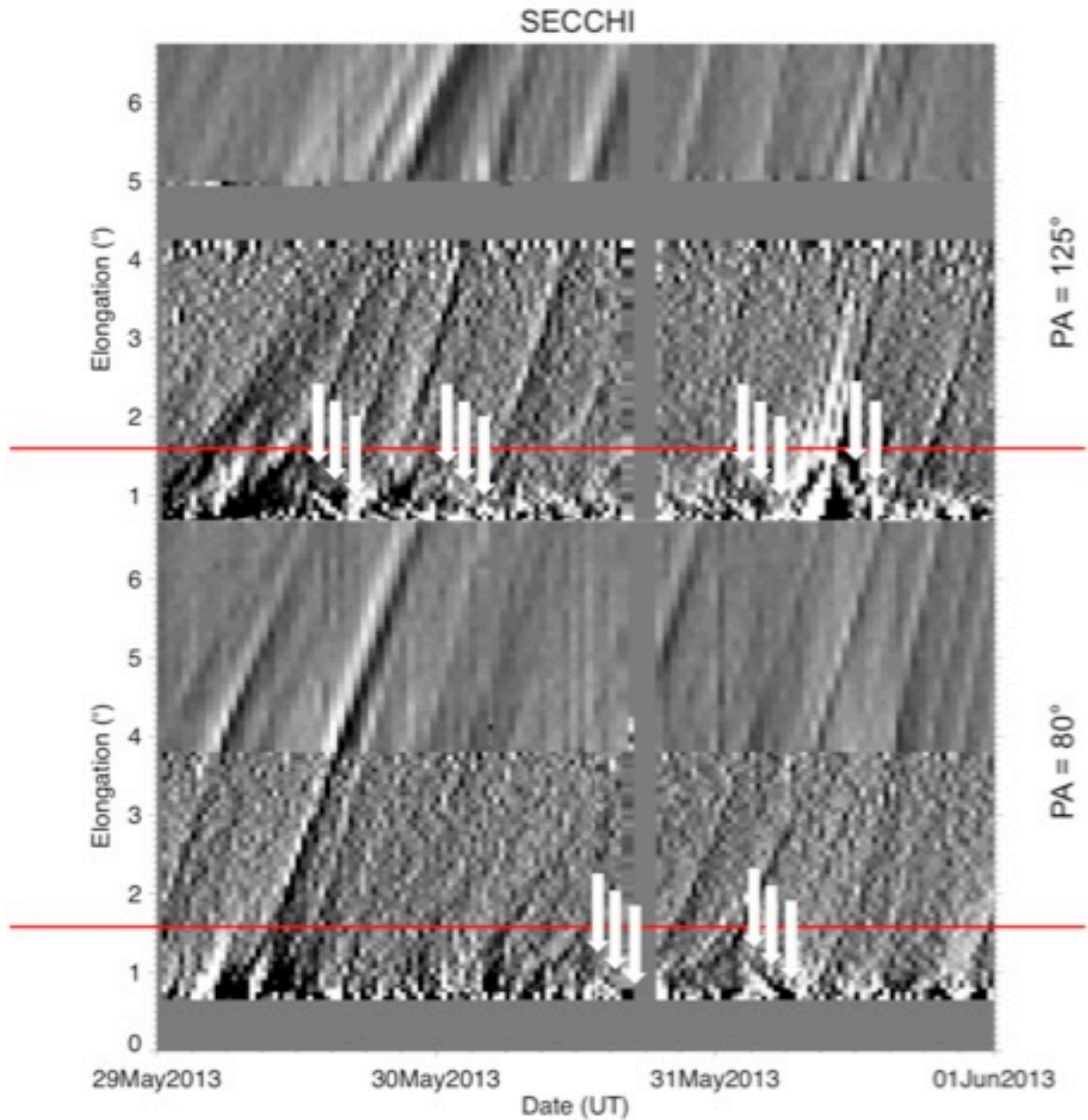

**Figure 4: J-map of SECCHI COR2-A and HI1-A during the passage of the CIR of May/June 2013 at PA 125° and 80°. The arrows point out some of the inflows associated with blobs. The red line shows the elongation of the outer edge of the LASCO C2 FOV.**

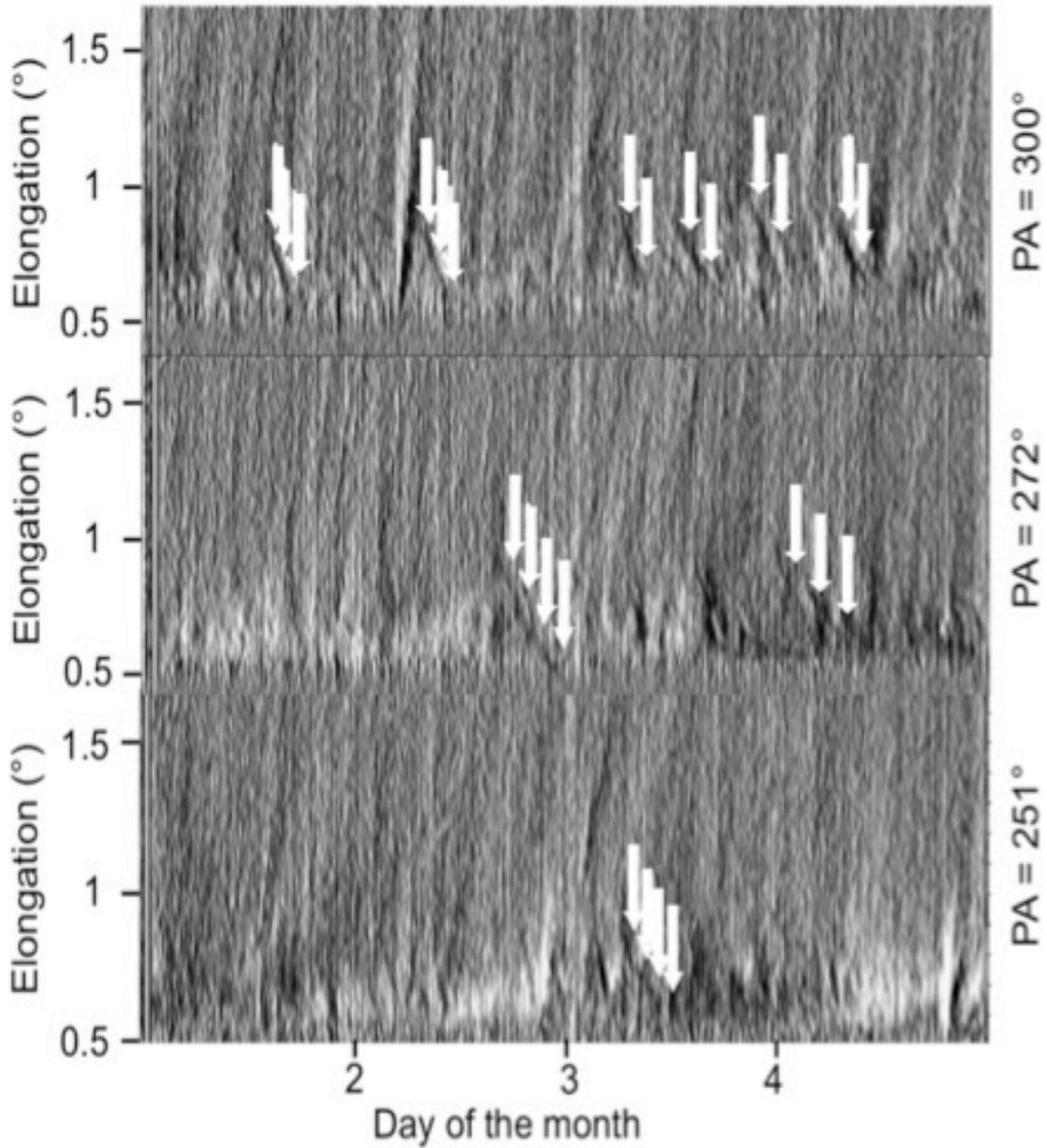

**Figure 5: J-map of LASCO C2 during the passage of the CIR of May/June 2013 at position angles 300°, 272° and 251°. The arrows indicate the signatures of some of the raining inflows.**